\documentstyle[twoside,epsfig,psfig]{article} 
 
\input ibvs2.sty 
 
\begin{document} 
\IBVShead{6080}{12 Nov 2013} 
 
\IBVStitle{Photometric evolution of Nova Del 2013 (V339 Del)}
\vskip -0.3 cm
\IBVStitle{during the optically thick phase}
\IBVStitle{} 
  
\IBVSauth{U. Munari$^1$, A. Henden$^2$, S. Dallaporta$^3$, G. Cherini$^3$} 

\IBVSinst{INAF Osservatorio Astronomico di Padova, Sede di Asiago, I-36032 Asiago (VI), Italy} 
\IBVSinst{AAVSO, 49 Bay State Rd. Cambridge, MA 02138, USA}
\IBVSinst{ANS Collaboration, c/o Astronomical Observatory, 36012 Asiago (VI), Italy}

\IBVStyp{Nova} 
\IBVSkey{photometry} 
\IBVSabs{}

\begintext 

The FeII-class Nova Del 2013 (=V339 Del) was discovered on 2013 Aug 14.584
UT by K.  Itagaki when it was already visible at unfiltered 6.8 magnitude
(cf.  CBET 3628). The progenitor seems to be the blue star USNO B-1
1107-0509795 with $B$$\sim$17.2, $R_{\rm C}$$\sim$17.4. The observation by
Denisenko et al.  (cf CBET 3628) reporting the nova still in quiescence at
$\sim$17.1 mag on Aug 13.998 UT (14 hours before the discovery), would indicate a
very fast rise to maximum.  The very convenient placement in the evening sky
and the naked-eye brightness attained by the nova favored a full suite of
all-out observing efforts, which have produced a flurry of circulars and
telegrams.  Among others, descriptions of nova spectra have been provided by
Shore et al.  (2013a,b,c,d), Munari et al.  (2013,a,b,c,d), Tomov et al. 
(2013), Darnley \& Bode (2013), Tarasova \& Shakhovskoi (2013), Woodward et
al.  (2013); infrared observations have been reported by Gehrz et al. 
(2013), Cass et al.  (2013a,b,c,d), Banerjee et al.  (2013a,b), Shenavrin et
al.  (2013); development of X-ray emission has been monitored by Nelson et
al.  (2013) and Page et al.  (2013a,b); and radio detections have been
described by Dutta et al.  (2013), Roy et al.  (2013), Chomiuk et al. 
(2013), and Anderson et al.  (2013).  There also seems to be a 5$\sigma$
detection of the nova in $\gamma$-rays (Hays et al.  2013).

No comprehensive description of the optical photometric evolution of Nova
Del 2013 has been provided to date.  In this paper we present our
$B$$V$$R_{\rm C}$$I_{\rm C}$ lightcurve of Nova Del 2013 covering the
evolution from discovery to day +77 past optical maximum, corresponding to
the end of the optically thick phase of the ejecta and their transition to
the nebular, optically thin regime.  The light- and color-curves are
presented in Figure~1, while Figure~2 highlights the phase of maximum brightness. 
The data are given in Table~1, available electronic only.
To observe the nova we used eight different robotic or remotely controlled
telescopes, equipped with photometric filters from various vendors (Schuler,
Omega, Astrodon).  Close to maximum brightness we used six small-diameter
refractors: four 6-cm Bright Star Monitor instruments (part of AAVSOnet;
located at AAVSO headquarters, Australia, California, New Mexico and
identified in Figures 1 and 2 as BSM HQ, AU, CA and NM, respectively), and
two operated by ANS Collaboration (a 6cm and a 15cm, with ANS identifiers 011
and 032.  At the time of maximum nova brightness their aperture was reduced
to 3 cm).  At later times, when the nova had faded several magnitudes below
maximum, two 30cm instruments joined the monitoring effort:
\clearpage

\IBVSfig{15cm}{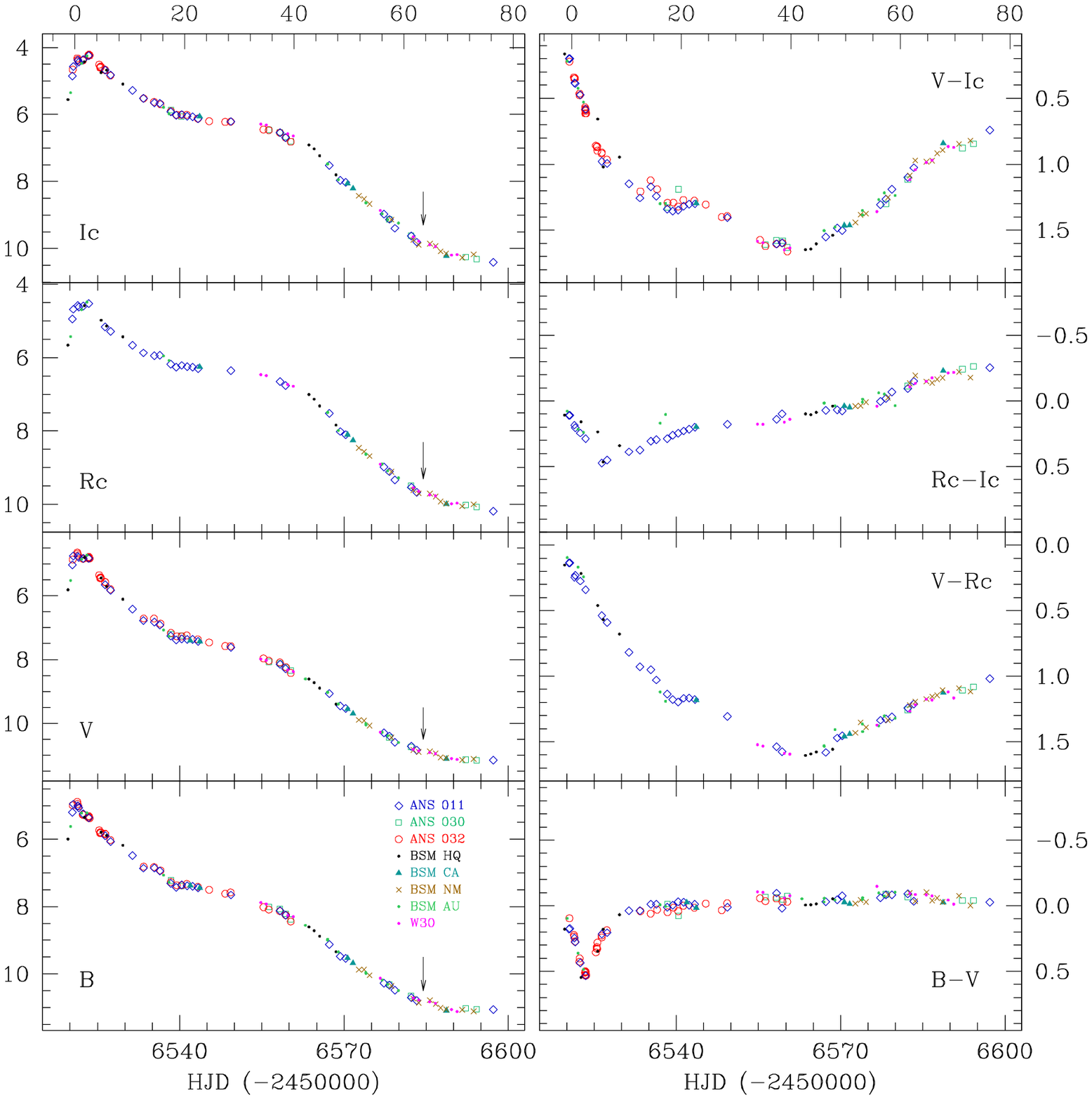}{Our $B$$V$$R_{\rm C}$$I_{\rm C}$ light- and
color-curves of Nova Del 2013 extending to day +77 past maximum, covering
the optically thick phase and the transition to nebular, optically thin
conditions.  The abscissae at the top of the graph are days counted from
maximum brightness in $V$ band.  The arrows mark the time (Oct 17)
when the flux of [OIII] 5007 \AA\ emission lines equalled that of H$\beta$.}

\noindent
W30 from AAVSOnet and 030 from ANS Collaboration.  The data collected at all
these eight telescopes were reduced against the same local photometric
sequence calibrated during photometric nights against Landolt (2009)
equatorial standards.  During data reduction, magnitude and colors were
obtained separately and were not derived one from the other.  The very
strong emission lines displayed by Nova Del 2013 introduce some systematic
deviation and offset (at the level of several hundredths of magnitude)
between the lightcurves obtained with different instruments, which cannot be
compensated for by the application of standard color equations.  To rectify
these small systematic offsets we have applied the lightcurve merging method
(LMM) described by Munari et al.  (2013e).  \clearpage

\IBVSfig{10.5cm}{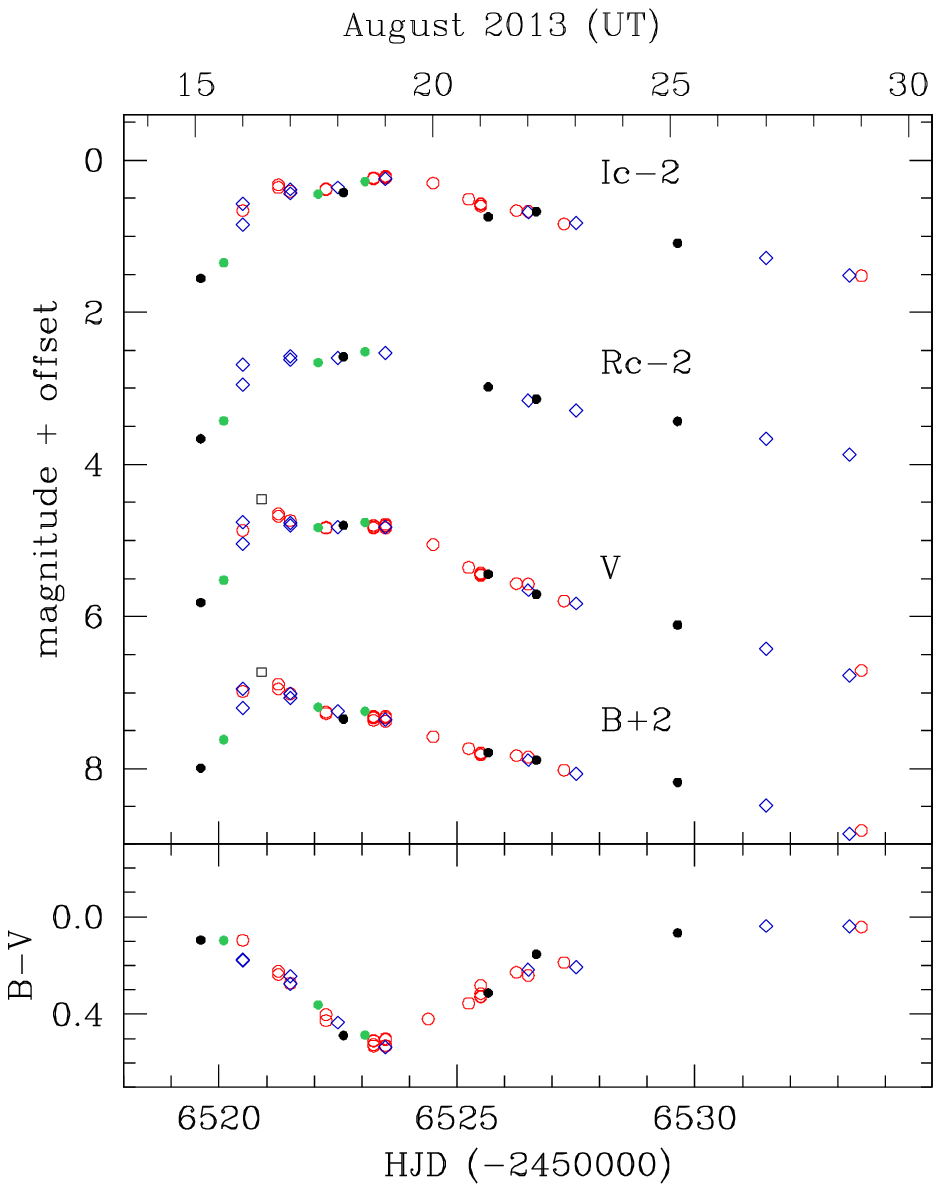}{Zoomed view of Figure~1 around Nova Del 2013 maximum
brightness.}

The photometric evolution of Nova Del 2013 as illustrated in Figure~1 and 2
is characterized by: (1) a smooth behavior, with short time scale
variations - if any was actually present - not exceeding at any time a few
hundredths of a magnitude; (2) a brief plateau appearing soon after maximum
brightness and lasting $\sim$1.5 days; (3) a longer plateau extending from
about day +20 to +37, that started when the nova had declined by $\Delta
V$=+2.8 mag from maximum brightness; (4) for both $B$ and $V$ bands, the
rate of brightness decline before and after this long plateau remained
the same: 0.125 mag/day.  The $R_{\rm C}$ and $I_{\rm C}$ bands behaved
similarly; (5) as typical of novae, a marked flattening of the decline rate
occurred simultaneously with the transition of the ejecta from optically
thick to nebular conditions.  The time when the flux of [OIII] 5007 \AA\
emission line equalled that of H$\beta$ (October 17, from Munari et al. 
2013e) is marked by an arrow in Figure~1.  This transition to nebular
conditions occured when the nova had declined by $\sim$6 mag, much more
than typically observed in other novae (where the transition occurs
$\sim$3.5/4.0 mag below maximum brightness); and (6) a peculiar value and
evolution for $B$$-$$V$ color, characterized by a nearly constant value
apart from a brief excursion around the time of maximum brightness and the
initial brief plateau.  According to van den Bergh \& Younger (1987),
typical novae display ($B$$-$$V$)$_\circ$=+0.23$\pm$0.06 at $B$,$V$ maximum
brightness and ($B$$-$$V$)$_\circ$=$-$0.02$\pm$0.04 at $t_{2}$.  The
corresponding observed values for Nova Del 2013 were $B$$-$$V$=+0.14 and
$B$$-$$V$=+0.04, that become ($B$$-$$V$)$_\circ$=$-$0.04 and
($B$$-$$V$)$_\circ$=$-$0.14 once corrected for the $E_{B-V}$$\simeq$0.18
insterstellar reddening affecting the nova (cf.  Tomov et al.  2013, Munari
et al.  2013a). 

Maximum light in $B$ and $V$ bands was attained by Nova Del 2013 during
daytime in Italy on August 16, and during bad weather conditions at the
AAVSOnet observing sites.  By interpolating data reported in CBET 3628, 3634
and by Burlak et al.  (2013) it can be estimated that maximum brightness
occurred around August 16.4 UT (JD 2456520.9), at $V$$\sim$4.46 and
$B$$\sim$4.70. The outburst amplitude is therefore $\Delta B$=12.5 mag,
comparing with $B$$\sim$17.2 for USNO B-1 1107-0509795 progenitor. This
maximum is marked by an open black square in Figure~2.  The evolution around
maximum brightness looks different depending on the photometric band.  In
the $B$ band, the maximum is sharp and the decline commenced immediately
with characteristic times $t_{2}^{B}$=12 and $t_{3}^{B}$=30 days, that were
$t_{2}^{V}$=10.5 and $t_{3}^{V}$=23.5 in $V$ band, which place Nova Del 2013
in a borderline position between {\it fast} and {\it very fast} novae
according to the Warner (1995) classification scheme.  
The brief plateau that the nova displayed in the $V$ band soon after maximum 
and that lasted $\sim$1.5 days, corresponded to a prolonged flat maximum in the
$R_{\rm C}$ band and to a delayed maximum in the $I_{\rm C}$ band, occuring
$\sim$2.5 days after that in $B$.

\references 

  Banerjee D.~P.~K., et al., 2013a, ATel, 5337, 1 

  Banerjee D.~P.~K., et al., 2013b, ATel, 5404, 1 

  Burlak A.~M., et al. 2013, ATel, 5294 
 
  Cass A.~C., et al., 2013a, ATel, 5419 

  Cass A.~C., et al., 2013b, ATel, 5434 

  Cass C.~A., et al., 2013c, ATel, 5317 

  Cass C.~A., et al., 2013d, ATel, 5340  

  Chomiuk L., et al., 2013, ATel, 5298

  Darnley M.~J., Bode M.~F., 2013, ATel, 5300

  Dutta P., et al., 2013, ATel, 5375  

  Gehrz R.~D., et al., 2013, ATel, 5299

  Hays E., et al., 2013, ATel, 5305 

  Landolt A.~U., 2009, AJ 137, 4186

  Munari U., Zwitter T., 1997, A\&A 318, 269

  Munari U., et al. 2013a, ATel, 5297

  Munari U., et al. 2013b, ATel, 5304

  Munari U., et al. 2013c, ATel, 5310

  Munari U., et al. 2013d, ATel, 5533

  Munari U., et al. 2013, New Astronomy 20, 30

  Nelson T., et al., 2013, ATel, 5305 

  Page K.~L., et al., 2013a, ATel, 5318 

  Page K.~L., et al., 2013b, ATel, 5470

  Roy N., et al., 2013, ATel, 5376

  Shenavrin V.~I., et al., 2013, ATel, 5431

  Shore S.~N., et al., 2013a, ATel, 5282

  Shore S.~N., et al., 2013b, ATel, 5312

  Shore S.~N., et al., 2013c, ATel, 5378

  Shore S.~N., et al., 2013d, ATel, 5409

  Tarasova T.~N., Shakhovskoi, D.~N., 2013, ATel, 5291

  Tomov T., et al., 2013, ATel, 5288

  van der Bergh S., Younger P.~F., 1987, A\&AS 70, 125

  Warner B., 1995, Cataclysmic Variable Stars, Cambridge Univ. Press

  Woodward C.~E., et al., 2013, ATel 5493

\endreferences 
\clearpage

\end{document}